\documentclass[12pt]{article}
\makeatletter
 \pdfoutput=1
\newcommand{\singlespacing}{\let\CS=\@currsize\renewcommand{\baselinestretch}{1}\tiny\CS}
\usepackage{epsfig}

\usepackage{lscape}
\usepackage{graphicx}
\usepackage{verbatim}   
\usepackage{color}      
\usepackage{amsfonts}
\usepackage{amsmath,bm, amssymb,amsthm}
\usepackage{float}
\usepackage{mathrsfs} 
\DeclareMathOperator*{\arg min}{arg\,min}
\usepackage{mathtools}
\usepackage{enumerate}
\usepackage{tikz}
\usepackage{makeidx}
\usepackage[left=2cm,right=2cm,top=2cm,bottom=2cm]{geometry}

\usepackage{float}
\usepackage{booktabs}
\usepackage{multirow}
\usepackage{caption}
\usepackage{subcaption}
\usepackage{array}
\usepackage{booktabs}
\usepackage{multirow}
\usetikzlibrary{patterns}
\usetikzlibrary{decorations.pathreplacing}
\usepackage{mathrsfs} 

\newcolumntype{d}{D{.}{.}{2.5}}
\newcolumntype{C}{>{\centering}p}

\begin{document}
\baselineskip=24pt
\parskip = 10pt
\def \qed {\hfill \vrule height7pt width 5pt depth 0pt}
\newcommand{\ve}[1]{\mbox{\boldmath$#1$}}
\newcommand{\IR}{\mbox{$I\!\!R$}}
\newcommand{\1}{\Rightarrow}
\newcommand{\bs}{\baselineskip}
\newcommand{\esp}{\end{sloppypar}}
\newcommand{\be}{\begin{equation}}
\newcommand{\ee}{\end{equation}}
\newcommand{\beanno}{\begin{eqnarray*}}
\newcommand{\inp}[2]{\left( {#1} ,\,{#2} \right)}
\newcommand{\eeanno}{\end{eqnarray*}}
\newcommand{\bea}{\begin{eqnarray}}
\newcommand{\eea}{\end{eqnarray}}
\newcommand{\ba}{\begin{array}}
\newcommand{\ea}{\end{array}}
\newcommand{\nno}{\nonumber}
\newcommand{\dou}{\partial}
\newcommand{\bc}{\begin{center}}
\newcommand{\ec}{\end{center}}
\newcommand{\2}{\subseteq}
\newcommand{\cl}{\centerline}
\newcommand{\ds}{\displaystyle}
\newcommand{\what}{\widehat}
\newcommand*\bigcdot{\mathpalette\bigcdot@{.5}}
\def\refhg{\hangindent=20pt\hangafter=1}
\def\refmark{\par\vskip 2.50mm\noindent\refhg}

\title{\sc Robust estimation of dependent competing risk model under interval monitoring and determining optimal inspection intervals }

\author{
Shuvashree Mondal$^{* 1}$ 
\& Shanya Baghel$^{2}$ 
 }

\date{}
\maketitle

\begin{abstract}

\noindent Recently, a growing amount interest is quite evident in modelling dependent competing risks in life time prognosis problem.  In this work, we propose to model the dependent competing risks by Marshal-Olkin bivariate exponential distribution.  The observable data consists of number of failures due to different causes across different time intervals.  The failure count data is common in instances like one shot devices where state of the subjects are inspected at different inspection times rather than the exact failure times.  The point estimation of the life time distribution in presence of competing risk has been studied through divergence based robust estimation method called minimum density power divergence estimation (MDPDE). The testing of hypothesis is performed based on a Wald type test statistic.  The influence function is derived both for the point estimator and the test statistic, which reflects the degree of robustness.  Another, key contribution of this work is to determine the optimal set of inspection times based on some predefined objectives. This article presents determination of  multi criteria based optimal design.  Population based heuristic algorithm  non-dominated
sorting-based multiobjective Genetic algorithm is exploited to solve this optimization problem.

\noindent 
    
\end{abstract}



\noindent$^{1,2}$ Department of Mathematics and Computing, Indian Institute of Technology (Indian School
of Mines) Dhanbad, Dhanbad- 826004, India.\\
\noindent \textbf{$^*$ Correspondence:} Shuvashree Mondal,
 Department of
Mathematics and Computing, Indian
Institute of Technology (Indian School of
Mines), Dhanbad 826004, India.\\
Email: shuvasri29@iitism.ac.in


\section{Introduction}

Competing risk data arises when an event takes place due to different simultaneously effective causes.  The occurrence of an event due to one specific cause precludes one from observing the occurrence of events due to the other causes. In the literature, a significant amount of work has been done in competing risk problem.  Crowder  \cite{Cr:2001} provided a monograph on the
analysis of different competing risk models.
Prentice et al.  \cite{P:1978} analysed failure time data in the competing risk environment.  Austin et al. \cite{A:2016} proposed the analysis of survival data in the presence of competing risks.  Balakrishnan et al. \cite{BN1:2015, BN2:2015} studied estimation of different lifetime distributions in presence of competing risks.  Balakrishnan et al. \cite{BN3:2015} provided Bayesian inference under competing risk setup.  Wang et al. \cite{wang:2020} studied competing risk failure time data for a frailty-copula model.  Dutta and Kayal \cite{Dutta:2022} conducted inferential study under censoring scheme on competing risk data.

\noindent Most of those articles present stochastically independent competing risks in action.  Recently, a growing interest is quite evident in modelling dependent competing risks in life time prognosis problem.  Justification of such modelling lies in instances like shock model originally found in Marshall and Olkin \cite{MO:1967}.  Suppose, in a system with two components, shock 1 is responsible for failure of component 1,  shock 2 is for component 2 while shock 3 results in failure of both the components. In such case, the system fails if any one component fails, it is indeed an example of dependent competing risk set up.  In the literature, we find the study on dependent competing risk in Bai et al. \cite{Bai:2020}, Cai et al. \cite{Cai:2017}, 
Feizjavdian and Hashemi \cite{FH:2015}, Kundu \cite{DK:2022}, Kundu and Mondal \cite{KM:2021}, 
 Shen and Xu \cite{SX:2018}, Lyu et al. \cite{Lyu:2022} and references therein.

\noindent In this work, we explore the study of statistical inference of the life time distribution under dependent competing risk set up.  It is assumed that life time under two dependent competing risks follows a bivariate Marshall Olkin distribution.  The subjects of interest are put on life testing experiment which continues until a pre-specified time point. The observable data consists of number of failures due to different causes across different time intervals.  The failure count data is common in instances like one shot devices where state of the subjects are inspected at different inspection times rather than the exact failure times, readers may see Blakrishnan and Ling \cite{BL:2012, BL1:2014}, Balakrishnan et al. \cite{BL2:2019} for references.

\noindent In inference study, conventional point estimation method is the maximum likelihood estimation (MLE) which is quite popular because of its well-known properties such as asymptotic efficiency, consistency, sufficiency, invariance transformation.  But in presence of outliers, MLE can not perform well.  Basu et al \cite{Basu:1998} proposed divergence based robust estimation method called minimum density power divergence estimator (MDPDE) by incorporating a tuning parameter which brings a trade-off between robustness and efficiency.  In this work, along with the MLE, we develop the MDPDE in dependent competing risk set up based on the failure count data.

\noindent Along with point estimation, testing of hypothesis is an essential component in inference study.  In this article, we present hypothesis testing based on the robust MDPDE. The null hypothesis is constituted based on the equality of the scale parameters of the competing risks.  In this regard a Wald type test statistic is developed  based on the asymptotic distribution of the MDPDE. An approximation method is applied for the power calculation.

\noindent The robustness of any statistic can be assessed by its influence function.  In the context of dependent competing risk set-up, the influence functions are computed both for the MDPDE and the Wald type test statistic.  Through numerical experiment also, we depict the robustness of the MDPDE compared to MLE.

\noindent Apart from inference study, in interval monitoring set-up, it is essential to set the inspection times such that the experiment serves different goals of the experimenter adequately.
In this context, we desire the precision of the estimator to be as high as possible along with minimum budget for the experiment. We try to achieve both the goals through multi-objective optimization.  Population based heuristic algorithm,  Genetic Algorithm (GA) is implemented which returns a set of Pareto optimal solutions. In the literature, Genetic algorithm has been successfully implemented in different situations. Readers may refer to Faraz \cite{Faraz:2013}, Liu et al. \cite{Liu:2015}, Parkinson \cite{Park:2000}, Yang et al.\cite{Yang:2021}.  In this work, we exploit a version of non-dominated sorting GA called NSGA-II proposed by Deb et al. \cite{Deb:2002}.

\noindent The rest of the article goes as follows.  In Section 2, we put down the description of the model along with the study of likelihood function and the maximum likelihood estimators.  We derive the robust density power divergence estimator in section 3.  Section 4 provides the study of the testing of hypothesis based on the robust estimator.  In Section 5, we study the influence functions for both point estimator and the test statistic.  Determination of optimal inspection times is studied in Section 6.  In Section 7, an extensive numerical experiment along with a real data analysis for illustration purposes are presented for the performance evaluation of the developed methods.

\section{Model Description}

In this section, we briefly describe the Marshall-Olkin Bivariate Exponential (MOBE) distribution as the life time model followed by description of the model layout.  \\
The cumulative distribution function (cdf) of an exponential distribution with scale parameter $\lambda$ is defined as
\beanno
F_{Exp}(x)=1- e^{-\lambda \ x }
\eeanno 
and the pdf is derived as
\beanno
f_{Exp}(x;\alpha, \lambda)= \lambda \ e^{-\lambda \ x }, \quad\text{where} \  x>0, \quad \lambda>0,
\eeanno
and it will be denoted by $Exp(\lambda)$.  Suppose, $U_0 \sim Exp(\lambda_0),\, U_1 \sim Exp(\lambda_1),\; U_2 \sim Exp(\lambda_2)$ and they are independently distributed. Define, $X_1 = min\{U_0, U_1\}$ and $X_2 = min\{U_0, U_2\}$.  The bivariate random vector $(X_1,X_2)$ is said to follow Marshall-Olkin bivariate Exponential distribution denoted by $MOBE(\lambda_0, \lambda_1, \lambda_2).$
The joint survival function of $(X_1,X_2)$ can be derived as, 
\beanno
S_{MOBE}(x_1,x_2)&=&P(X_1>x_1,X_2>x_2)=P(U_0>z,U_1>x_1,U_2>x_2)\nonumber\\
&=& e^{-(\lambda_0 z + \lambda_1 x_1 + \lambda_2 x_2) }
\eeanno
where  $z = max\{x_1, x_2\}.$ 
Therefore, the joint probability density function (PDF) of $(X_1, X_2)$ can be obtained as
\begin{equation}
\label{S2.E3}
f_{MOBE}(x_1,x_2)  = \left\{
\begin{array}{ll}
\displaystyle \lambda_1 (\lambda_0+\lambda_2) e^{- \lambda_1 x_1 - (\lambda_0+\lambda_2)x_2  } &~  0 < x_1 < x_2 < \infty\\ 
\displaystyle 
\lambda_2 (\lambda_0+\lambda_1) e^{- (\lambda_0+\lambda_1 ) x_1 - \lambda_2)x_2  } &~  0 < x_2 < x_1 < \infty\\ 
\displaystyle   \displaystyle \lambda_0 e^{- \lambda x}, &~  0 < x_1=x_2=x < \infty.\\ 
\end{array} \right. 
\end{equation}
where, $\lambda=\lambda_0+\lambda_1+\lambda_2.$\\
\noindent  Suppose n units are put on the life testing experiment and each unit is subject to two competing risks.  Let $T_1$ denote the failure time due to risk 1 and $T_2$ denote the same for risk 2.  Here, we assume that $(T_1, T_2)\sim MOBE( \lambda_0, \lambda_1, \lambda_2).$
Under these competing risk set-up, the observable failure time is $T=min(T_1, T_2).$  In the life testing experiment, at different inspection times say $\tau_1, \ldots, \tau_K,$ the experimenter will observe the number of failures in each interval due to the competing causes and the experiment is terminated at $\tau_K$ time point.  Let $N_{i}$ be the  number of failures which take place in $(\tau_{i-1}, \tau_{i}]$ interval for $i=1, \ldots, K$ where $\tau_0=0.$  $N_{i}$ can be decomposed as $N_{i}=N_{i0}+N_{i1}+N_{i2}$, where $N_{i1} (N_{i2} )$ is the number of failures due to cause l (cause 2) and $N_{i0}$ is the number of failure due to both the causes. Let $N_s$ be the censored units at the time point $\tau_K$, therefore, $N_s=n-\sum_{i=1}^{K}\sum_{l=0}^{2}N_{il}.$ \\ 
It is evident that, $(N_{11},N_{12},N_{10}, \cdots, N_{K1},N_{K2},N_{K0}, N_{s})\sim  \it{Multinomial}(n, \bf{p}),$
with the probability vector ${\bf{p}}=(p_{11},p_{12},p_{10},\cdots, p_{K1},p_{K2},p_{K0},p_{s})$, where for $i=1,\ldots, K,$
\beanno
p_{i1}&=&P(\tau_{i-1}<T_1\leq \tau_{i}, T_2>T_1)\nonumber\\
&=& \int_{\tau_{i-1}}^{\tau_{i}}\int_{x_2}^{\infty} f_{MOBE}(x_1,x_2)I(x_1<x_2)~dx_1~dx_2\\
&=&\frac{\lambda_1}{\lambda}\left(e^{-\lambda \tau_{i-1} }-e^{-\lambda \tau_{i} }\right)\\
p_{i2} &=&P(\tau_{i-1}<T_2\leq \tau_{i}, T_1>T_2) \\
&=&
\frac{\lambda_2}{\lambda}\left(e^{-\lambda \tau_{i-1}}-e^{-\lambda \tau_{i} }\right)\\ 
p_{i0} &=&P(\tau_{i-1}<T_1= T_2\leq \tau_{i}) \\
&=&\frac{\lambda_0}{\lambda}\left(e^{-\lambda \tau_{i-1} }-e^{-\lambda \tau_{i} }\right), \quad \text{and} \\
p_{s} &=& P(min (T_1, T_2) > \tau_K)\\
&=&e^{-\lambda \tau_{K} }.
\eeanno

\noindent Based on the failure count data across the intervals, the likelihood function can be written as
\begin{eqnarray}
L({ \boldsymbol{\theta} })&\propto& \left(\prod_{i=1}^{K}\prod_{l=0}^{2} p_{il}^{N_{il}}\right)\times p_{s}^{N_{s}}\nonumber\\
&=& \frac{\lambda_1^{\sum_{i=1}^{K}N_{i1} }\lambda_2^{\sum_{i=1}^{K}N_{i2}} \lambda_0^{\sum_{i=1}^{K}N_{i0} }}{\lambda^{\sum_{i=1}^{K} N_i }}\times \prod_{i=1}^{K}\left(e^{-\lambda \tau_{i-1}}-e^{-\lambda \tau_{i}}\right)^{N_{i}}\times e^{-\lambda N_{s}\tau_{K} }\nonumber
\end{eqnarray}
where ${ \boldsymbol{\theta} }=(\lambda_0, \lambda_1, \lambda_2)^T.$\\
Therefore, the log-likelihood can be written as
\begin{eqnarray}
l({ \boldsymbol{\theta} } )
&=& \sum_{i=1}^{K}N_{i1}\log{\lambda_1}+\sum_{i=1}^{K}N_{i2}\log{\lambda_2}+\sum_{i=1}^{K} N_{i0}\log{\lambda_0}-\sum_{i=1}^{K} N_{i}\log{\lambda}\nonumber\\&&+ \sum_{i=1}^{K}N_{i}\log\left(e^{-\lambda \tau_{i-1} }-e^{-\lambda \tau_{i} }\right)-\lambda N_{s}\tau_{K}.\nonumber
\end{eqnarray}
The estimating equations are
\beanno
\frac{\sum_{i=1}^{K} N_{ij} }{\lambda_j} - \frac{\sum_{i=1}^{K} N_i }{\lambda} + \frac{ \sum_{i=1}^{K} N_i (\tau_ie^{-\lambda \tau_{i} } - \tau_{i-1} e^{-\lambda \tau_{i-1} } )         }{ (e^{-\lambda \tau_{i-1} }-e^{-\lambda \tau_{i}  } )}  - N_s \tau_K=0
\eeanno
for $j=0,1,2.$

\noindent Though MLE is a very popular estimator due to its several properties like consistency, efficiency, it is not able to perform the analysis well in presence of outliers in the dataset. In the following section we will study a robust estimation method to obtain the estimates of the unknown parameter $\boldsymbol\theta.$

\section{The Method of Density Power Divergence}

Basu et.al \cite{Basu:1998} first developed the density power divergence method for robust estimation.  They considered a parametric family of models with  densities $\{f_t\}$ with respect to Lebesgue measure where unknown parameter $t\in \Omega$, which is the parameter space.  With respect to the same  measure, let $G$ be the class of all distributions having densities $g$.
Under these assumptions they define the divergence between density functions $g$ and $f_t$ as
\begin{eqnarray}
\label{S3.E1}
d_{\beta}(g, f_t)=\int \Bigg\{f_t^{1+\beta}(u)-(1+\frac{1}{\beta}) g(u)\,f_t^{\beta}(u) +\frac{1}{\beta}g^{1+\beta}(u)\Bigg\}~du,\quad \beta>0.
\end{eqnarray}
Note that when $\beta$ tends to 0, $d_{\beta}(g, f_t)$ tends to become the Kullback-Leibler divergence between $g$ and $f_t.$  In case of having a random sample $X_1, \ldots, X_n,$ from G, the true distribution G can be replaced by the empirical distribution and the minimum density power divergence estimate (MDPDE) is the value of the parameter t, which will minimize 
\beanno
d_{\beta}(g, f_t)=\int f_t^{1+\beta}(u)du -(1+\frac{1}{\beta}) \sum_{i=1}^{n}f_t^{\beta}(X_i).
\eeanno
In our context, the density power divergence  between the theoretical probability vector ${\bf{p}}=(p_{11},p_{12},p_{10},\ldots, p_{K1},p_{K2},p_{K0},p_{s})$ and the empirical measure ( $\frac{N_{11}}{n},$ $\frac{N_{12}}{n},$ $\frac{N_{10}}{n},$  $\ldots,$ $\frac{N_{k1}}{n},$  $\frac{N_{k2}}{n},$ $\frac{N_{k0}}{n},$ \\ $\frac{N_{s}}{n}$ ) can be obtained as,
\begin{eqnarray}
\label{S3.E2}
d_{\beta} &=&\sum_{i=1}^{K}\sum_{j=0}^{2}p_{ij}^{1+\beta}+p_{s}^{1+\beta}-\frac{1+\beta}{\beta}\left[ \sum_{i=1}^{K}\sum_{j=0}^{2}\left(\frac{N_{ij}}{n}p_{ij}^{\beta}\right)+\frac{N_{s}}{n}p_{s}^{\beta}\right]\nonumber\\&&+\frac{1}{\beta}\left[ \sum_{i=1}^{K}\sum_{j=0}^{2}\left(\frac{N_{ij}}{n}\right)^{1+\beta}+\left(\frac{N_{s}}{n}\right)^{1+\beta}\right].
\end{eqnarray}
Minimizing $d_{\beta}$  with respect to the parameters $\lambda_0, \lambda_1, \lambda_2$ is equivalent as minimizing $H_{n}(\beta)$ where,
\bea
\label{S3.E3}
H_{n}(\beta)=  
&\sum_{i=1}^{K}\sum_{j=0}^{2}p_{ij}^{1+\beta}+p_{s}^{1+\beta}-\frac{1+\beta}{\beta}\left[\sum_{i=1}^{K}\sum_{j=0}^{2}\left(\frac{N_{ij}}{n}p_{ij}^{\beta}\right)+\frac{N_{s}}{n}p_{s}^{\beta} \right].
\eea
Based on \eqref{S3.E2} and \eqref{S3.E3}, the minimum density power divergence estimator of $\boldsymbol{\theta}=(\lambda_0, \lambda_1, \lambda_2)^{T}$
can be derived as
\beanno
{\what{\boldsymbol{\theta}} }_{\beta} = \arg min _{\boldsymbol{\theta} } H_{n}(\beta); \ \beta>0.
\eeanno
The set of estimating equations can be obtained as
\beanno
\sum_{i=1}^{k}\sum_{j=0}^{2}p_{ij}^{1+\beta} \frac{\partial \log p_{ij} }{\partial \boldsymbol{\theta}    }    +p_{s}^{1+\beta}  \frac{\partial \log p_{s} }{\partial \boldsymbol{\theta}    }- \left[\sum_{i=1}^{k}\sum_{j=0}^{2}\left(\frac{N_{ij}}{n}p_{ij}^{\beta}  \frac{\partial \log p_{ij} }{\partial \boldsymbol{\theta}    }    \right)+\frac{N_{s}}{n}p_{s}^{\beta} \frac{\partial \log p_{s} }{\partial \boldsymbol{\theta}    }\right] = 0.
\eeanno
The estimating equations are unbiased and the estimator is Fisher consistent.\\
\noindent In the following result, the asymptotic distribution of the MDPD estimator is presented for Marshal-Olkin bivariate exponential distribution under competing risk set up.\\
\noindent \textbf{Result 1:}  Let $\boldsymbol{\theta}_0$ be the true value of parameter 
$\boldsymbol{\theta}.$ The asymptotic distribution of the MDPD estimator 
${\hat{\boldsymbol{\theta}} }_{\beta}$ is given by
\beanno
\sqrt{n}({\what{\boldsymbol{\theta}} }_{\beta} - {\boldsymbol{\theta}_0}) \sim N(0_3, J_{\beta}({\boldsymbol{\theta}_0})^{-1} K_{\beta}({\boldsymbol{\theta}_0})J_{\beta}({\boldsymbol{\theta}_0})^{-1} )
\eeanno
where,  $J_{\beta}({\boldsymbol{\theta}_0})$ and $K_{\beta}({\boldsymbol{\theta}_0})$ are defined in Appendix.\\
Proof: See in Appendix.

\section{ Robust test statistics for hypothesis testing: }

In this section, a testing of hypothesis is developed based on the asymptotic distribution of the MDPDE on the same way as Wald test .  Suppose, the null hypothesis is set as $H_0: \lambda_1=\lambda_2$ where the alternative hypothesis $H_1: \lambda_1  \neq \lambda_2.$  Therefore, it can be rewritten as $H_0: a^{T}_0{\boldsymbol{\theta}}=0$ where $a_0=(0,1,-1)^{T}$ or equivalently, $H_0: \boldsymbol{\theta } \in \boldsymbol{\Theta}_0$ where $\Theta_0 = \{ \theta: 
a^{T}_0\boldsymbol{ \theta}=0 \}$ and $H_1: \boldsymbol{\theta} \notin \Theta_0.$

\noindent  When $\boldsymbol{\theta}_0$ is the true value of $\boldsymbol{\theta},$ $\sqrt{n}({\what{\boldsymbol{\theta}} }_{\beta} - {\boldsymbol{\theta}_0}) \sim N(0_3, \Sigma(\boldsymbol{\theta}_0))$
where 
$\Sigma(\boldsymbol{\theta}_0)=
J_{\beta}({\boldsymbol{\theta}_0})^{-1} K_{\beta}({\boldsymbol{\theta}_0})J_{\beta}({\boldsymbol{\theta}_0})^{-1},$  and under $H_0$
\beanno
 \sqrt{n} a^{T}_0 {\what{\boldsymbol{\theta}} }_{\beta} \sim N\Big( 0_r, a^{T}_0
\Sigma(\boldsymbol{\theta}_0)a_0 \Big).
\eeanno
Therefore, the test statistic can be defined as,
\beanno
M_n( {\what{\boldsymbol{\theta}} }_{\beta}  ) = n (a^{T}_0 {\what{\boldsymbol{\theta}} }_{\beta} )^{T} \Big( a^{T}_0
\Sigma({\what{\boldsymbol{\theta}} }_{\beta}) a_0  \Big)^{-1} (a^{T}_0 {\what{\boldsymbol{\theta}} }_{\beta} ). \eeanno
When $H_0$ is true, $M_n( {\what{\boldsymbol{\theta}} }_{\beta}  ) \sim \chi^{2}_1$ and at level
$\alpha,$ the rejection region can be obtained as $M_n( {\what{\boldsymbol{\theta}} }_{\beta}  ) \geq \chi^{2}_{1, \alpha}$ where $ \chi^{2}_{1, \alpha}$ is the upper $\alpha$ percentile of  $ \chi^{2}_{1}.$

\noindent In this testing, the power function can not be calculated explicitly.  Therefore, an approximation method is being implemented as suggested by Basu el al \cite{Basu:2017}.  Define, 
\beanno
m(\boldsymbol{\theta}_1, \boldsymbol{\theta}_2) =  (a^{T}_0 {\boldsymbol{\theta}}_1 )^{T} \Big( a^{T}_0
\Sigma({\boldsymbol{\theta}}_2) a_0  \Big)^{-1} (a^{T}_0 {\boldsymbol{\theta}}_1 ).
\eeanno
where ${\boldsymbol{\theta}}_1,$ ${\boldsymbol{\theta}}_2$ are some values of ${\boldsymbol{\theta}}.$\\
\noindent Taylor series expansion of $m({\what{\boldsymbol{\theta}} }_{\beta}, \boldsymbol{\theta^{*}})$ at $\what{\boldsymbol{\theta}}_{\beta}$ around $\boldsymbol{\theta^{*}}$ is given as
\beanno
m({\what{\boldsymbol{\theta}} }_{\beta}, \boldsymbol{\theta^{*}}) =   m({{\boldsymbol{\theta^{*} }} }, \boldsymbol{\theta^{*}}) + \frac{\partial m({{\boldsymbol{\theta}} }, \boldsymbol{\theta^{*}})}{\partial \boldsymbol{\theta^{T}}      } \vert_{{\boldsymbol{\theta}} =\boldsymbol{\theta^{*}} } ( {\what{\boldsymbol{\theta}} }_{\beta} -  \boldsymbol{\theta^{*}} ) +o_p(   || {\what{\boldsymbol{\theta}} }_{\beta}- \boldsymbol{\theta^{*}}   || ).
\eeanno
When, $\boldsymbol{\theta}=\boldsymbol{\theta^{*}} \notin \boldsymbol{\Theta}_0,$
\beanno
\sqrt{n} \Big( m({\what{\boldsymbol{\theta}} }_{\beta}, \boldsymbol{\theta^{*}}) -   m({{\boldsymbol{\theta^{*} }} }, \boldsymbol{\theta^{*}})  \Big) \sim N( 0, \sigma^{2}(\boldsymbol{\theta^{*}} ) )
\eeanno
where, $\sigma^{2}(\boldsymbol{\theta^{*}} ) = \frac{\partial m({\boldsymbol{\theta}} , \boldsymbol{\theta^{*}})}     {\partial \boldsymbol{\theta^{T}}      } \vert_{{\boldsymbol{\theta}} ={\boldsymbol{\theta^{*} }} }   \Sigma(\boldsymbol{\theta^{*} })  \frac{\partial m({\boldsymbol{\theta}} , \boldsymbol{\theta^{*}})}     {\partial \boldsymbol{\theta }      } \vert_{{\boldsymbol{\theta}} ={\boldsymbol{\theta^{*} }} }.$

\noindent Therefore, the power of the test at $\boldsymbol{\theta}=\boldsymbol{\theta^{*}}$ can be obtained as
\beanno
P\Big( M_n( {\what{\boldsymbol{\theta}} }_{\beta}  ) \geq \chi^{2}_{1, \alpha} \Big)&=& P\Bigg( n\Big(   m({\what{\boldsymbol{\theta}} }_{\beta}, {\what{\boldsymbol{\theta}} }_{\beta}) - m({{\boldsymbol{\theta^{*} }} }, \boldsymbol{\theta^{*}}) \Big  ) \geq    \chi^2_{1, \alpha} - n m({{\boldsymbol{\theta^{*} }} }, \boldsymbol{\theta^{*}}) \Bigg) \\
&=& P\Bigg(   \frac{\sqrt{n} \Big(m({\what{\boldsymbol{\theta}} }_{\beta}, {\what{\boldsymbol{\theta}} }_{\beta}) - m({{\boldsymbol{\theta^{*} }} }, \boldsymbol{\theta^{*}}) \Big)}{ \sigma(\boldsymbol{\theta^{*}} )  } \geq 
\frac{1}{\sigma(\boldsymbol{\theta^{*}} )} \Big( \frac{ \chi^2_{1, \alpha}}{\sqrt{n}} - \sqrt{n}m({{\boldsymbol{\theta^{*} }} }, {{\boldsymbol{\theta^{*} }} }   ) \Big)
\Bigg) \\
&=& 1- \Phi\Bigg( \frac{1}{\sigma(\boldsymbol{\theta^{*}} )} \Big( \frac{ \chi^2_{1, \alpha}}{\sqrt{n}} - \sqrt{n}m({{\boldsymbol{\theta^{*} }} }, {{\boldsymbol{\theta^{*} }} }   )      \Big) \Bigg).
\eeanno

\section{ \sc Robustness Property :}

Robustness of any estimator can be expressed through its influence function (IF).  This section presents the influence function of the MDPD point estimator and the Wald type test statistic.

\subsection{\sc Influence function of  MDPDE : }  
Suppose G is the true distribution from where data have been generated. If $T_{\beta}(G)$ denotes the statistical functional of the MDPDE $\what{\boldsymbol{\theta}}_{\beta},$ then $T_{\beta}(G)$ be the value of $\boldsymbol{\theta}$ which will minimize 
\beanno
&\sum_{i=1}^{k}\sum_{j=0}^{2}p_{ij}^{1+\beta}+p_{s}^{1+\beta}-\frac{1+\beta}{\beta}\left[\sum_{i=1}^{k}\sum_{j=0}^{2}\left(\int_{I_{ij}}dG \right ) p_{ij}^{\beta}+\int_{I_s} dGp_{s}^{\beta} \right].
\eeanno
where $(x_1, x_2) \in I_{i1} \implies (\tau_{i-1} < x_1 \leq \tau_i, x_2 > x_1),$
$(x_1, x_2) \in I_{i2} \implies (\tau_{i-1} < x_2 \leq \tau_i, x_1 > x_2),$
$(x_1, x_2) \in I_{i0} \implies (\tau_{i-1} < x_1=x_2 \leq \tau_i),$
for $i=1,\ldots,K$ and
$(x_1, x_2) \in I_{s} \implies (x_1> \tau_K, x_2 >\tau_K).$\\
\noindent Therefore, $T_{\beta}(G)$ will satisfy
\bea
\label{if-1}
&\sum_{i=1}^{k}\sum_{j=0}^{2}p_{ij}^{1+\beta} \frac{\partial \log p_{ij} }{\partial \boldsymbol{\theta}    }     +p_{s}^{1+\beta} \frac{\partial \log p_{s} }{\partial \boldsymbol{\theta}    }  -\left[\sum_{i=1}^{k}\sum_{j=0}^{2}\left(\int_{I_{ij}}dG \right ) p_{ij}^{\beta}  \frac{\partial \log p_{ij} }{\partial \boldsymbol{\theta}    }  +\int_{I_s} dGp_{s}^{\beta}  \frac{\partial \log p_{s} }{\partial \boldsymbol{\theta}    } \right] =0.
\eea
Then the influence function is obtained as
\beanno
IF(\boldsymbol{x}, T_{\beta}, G) =\lim_{\epsilon \to 0} \frac{T_{\beta}(G_{\epsilon}) - T_{\beta}(G) }{\epsilon} = \frac{\partial T_{\beta}(G_{\epsilon}) }{\partial \epsilon} \vert_{\epsilon =0}.
\eeanno
where $\boldsymbol{x}=(x_1,x_2), x_1,x_2 \in (0, \infty),$ $G_{\epsilon}=(1-\epsilon) G + \epsilon \Delta_{\boldsymbol{x}}$ and $\Delta_{\boldsymbol{x}}$ is the degenerate distribution with point mass 1 on $\boldsymbol{x}.$  In the following result, the influence function is derived under the assumed set-ups.\\
\noindent \textbf{Result 2:} The IF of $\what{\boldsymbol{\theta}}_{\beta},$ for the
Marshal-Olkin bivariate exponential distribution under competing risk set up
is given by
\beanno
IF(\boldsymbol{x}, T_{\beta}, F_{\boldsymbol{ \theta} } )= J_{\beta}({\boldsymbol{\theta} }  )^{-1}\Big[ \sum_{i=1}^{k} \sum_{j=0}^{2} (\delta_{I_{ij}}(\boldsymbol{x}) - p_{ij} ) p^{\beta}_{ij}\frac{\partial \log{p_{ij}} }{\partial \boldsymbol{\theta} } + (\delta_{I_{s}}(\boldsymbol{x} ) - p_{s} ) p^{\beta}_{s}\frac{\partial \log{p_{s}} }{\partial \boldsymbol{\theta}} \Big].
\eeanno
where
$\delta_{A}(\boldsymbol{x}) = 
\begin{cases}
1 \quad \text{if} \  \boldsymbol{x} \in A\\
0 \quad \text{otherwise}. \\
\end{cases}
$
\\
\noindent Proof: See in Appendix. \\
\noindent The maximum of this Influence function over $\boldsymbol{x}$ indicates the extent of
bias due to contamination. Therefore, smaller value of IF will indicate the estimator as more robust.

\subsection{\sc Influence function of  Wald type test statistics : }  
The statistical functional of $M_n(\what{\boldsymbol{\theta}}_{\beta})$ can be obtained as $M_n(T_{\beta}(G) ).$  Therefore the influence function of $M_n()$ is given by $IF( \boldsymbol{x}, M_n, F_{\boldsymbol{ \theta}_0} ) =
\lim_{\epsilon \to 0}  \frac{\partial M_n(T_{\beta} (G_{\epsilon}  ) ) } { \partial \epsilon}$
where, $G_{\epsilon}=(1-\epsilon)F_{\boldsymbol{\theta_0}}  + \epsilon \Delta_{\boldsymbol{x}}.$
\beanno
\frac{\partial M_n(T_{\beta} (G_{\epsilon}))}  {\partial \epsilon}
&=& \frac{\partial T^{T}_{\beta}(G_{\epsilon}) }{\partial \epsilon} a_0 \Big( a^{T}_0
\Sigma(\boldsymbol{\theta}_0 ) a_0  \Big)^{-1} (a^{T}_0
T_{\beta}(G_{\epsilon}))\\
&& +  (a^{T}_0
T_{\beta}(G_{\epsilon}))^{T} \frac{\partial   \Big( a^{T}_0
\Sigma(\boldsymbol{\theta}_0 )  a_0  \Big)^{-1}    }{\partial \epsilon }  (a^{T}_0
T_{\beta}(G_{\epsilon})) \\
&& +  (a^{T}_0
T_{\beta}(G_{\epsilon})  )^{T} \Big( a^{T}_0
\Sigma( \boldsymbol{\theta}_0 ) a_0  \Big)^{-1} a^{T}_0 \frac{\partial T_{\beta}(G_{\epsilon}) }{\partial \epsilon}
\eeanno
As, $T(F_{\boldsymbol{\theta}_0 })=\theta_0 \in \Theta_0$ and $a^T_0 \theta_0=0,$  the first order influence function 
$IF( \boldsymbol{x}, M_n, F_{\boldsymbol{ \theta}_0} )=0.$
The second order influence function is derived as $IF_2( \boldsymbol{x}, M_n, F_{\boldsymbol{ \theta}_0} ) =
\lim_{\epsilon \to 0}  \frac{\partial^2 M_n(T_{\beta} (G_{\epsilon}  ) ) } { \partial \epsilon^2 }.$
\beanno
\frac{\partial^2 M_n(T_{\beta} (G_{\epsilon}))}  {\partial \epsilon^2} &=&
\frac{\partial^2 T^{T}_{\beta}(G_{\epsilon}) }{\partial \epsilon^2} a_0 \Big( a^{T}_0
\Sigma(\boldsymbol{ \theta}_0 ) a_0  \Big)^{-1} (a^{T}_0
T_{\beta}(G_{\epsilon}))\\
&& +  (a^{T}_0
T_{\beta}(G_{\epsilon}))^{T} \frac{\partial^2  \Big( a^{T}_0
\Sigma( \boldsymbol{ \theta}_0)  a_0  \Big)^{-1}    }{\partial \epsilon^2 }  (a^{T}_0
T_{\beta}(G_{\epsilon})) \\
&& +  (a^{T}_0
T_{\beta}(G_{\epsilon})  )^{T} \Big( a^{T}_0
\Sigma(\boldsymbol{ \theta}_0 ) a_0  \Big)^{-1} a^{T}_0 \frac{\partial^2 T_{\beta}(G_{\epsilon}) }{\partial \epsilon^2} \\
&& + 2 \frac{\partial T^{T}_{\beta}(G_{\epsilon}) }{\partial \epsilon} a_0
\frac{\partial   \Big( a^{T}_0
\Sigma(\boldsymbol{ \theta}_0  ) a_0  \Big)^{-1}    }{\partial \epsilon } (a^{T}_0
T_{\beta}(G_{\epsilon})) \\
&& + 2 (a^{T}_0
T_{\beta}(G_{\epsilon})  )^{T} \frac{\partial   \Big( a^{T}_0
\Sigma(\boldsymbol{ \theta}_0 ) a_0  \Big)^{-1}    }{\partial \epsilon } a^{T}_0 \frac{\partial T_{\beta}(G_{\epsilon}) }{\partial \epsilon} \\
&& 2 \frac{\partial T^{T}_{\beta}(G_{\epsilon}) }{\partial \epsilon} a_0
\Big( a^{T}_0
\Sigma(\boldsymbol{ \theta}_0  ) a_0  \Big)^{-1} a^T_0\frac{\partial T_{\beta}(G_{\epsilon}) }{\partial \epsilon} 
\eeanno
Therefore,  $IF_2(   \boldsymbol{x}, M_n, F_{\boldsymbol{ \theta}_0} ) = 2 IF^T(   \boldsymbol{x}, T_{\beta}, F_{\boldsymbol{ \theta}_0} ) a_0   \Big( a^{T}_0
\Sigma(\boldsymbol{ \theta}_0 ) a_0  \Big)^{-1} a^T_0 IF(   \boldsymbol{x}, T_{\beta}, F_{\boldsymbol{ \theta}_0} ).$

\section{\sc Optimal inspection times}

\noindent In this experimental set-up, the design parameter of the life testing experiment consists of $\mathscr{D}=(\tau_1, \ldots, \tau_K)$ for fixed sample size.  To the experimenters, the concerned aspects influenced by the design are the cost of the experiment and the precision of the estimators of the model parameters.  Increasing the precision of the estimators is equivalent to minimizing trace or determinant of the covariance matrix of the estimators.  With low precision, the estimated values of the model parameters are highly unreliable which may result in wrong prediction of the life time distribution.  Hence, it is desirable to set the inspection times such that those concerned issues can be overcome. 
The optimal design is the set of time points which will determine the best experimental set-up based on some objectives defined by the experimenter.

\noindent In the literature, determination of optimal design or plan is found in wide spectrum of applications.  In Balakrishnan and Han \cite{BN:2009}, the optimal plan was studied in application of accelerated life testing.  In reliability analysis of one-shot devices optimal inspection times was determined in Ling \cite{Ling:2019}. Ng et al. \cite{Ng:2004} studied optimal plan in application of progressive censoring schemes.  Recently, Bhattacharya \cite{BR:2020} et al. studied multi-criteria based optimal life testing plan.  This article presents determination of  multi criteria based optimal design.

\noindent In this work, both the experiment cost and determinant of covariance matrix of the estimators will be minimized simultaneously. The experiment cost is defined in the following objective function as
\beanno
\Phi_1 = C_0 + C_n n +C_f E(n-N_s)
\eeanno
where $C_n$ is the cost per unit put on in the experiment, $C_f$ indicates cost per failure in the experiment and $C_0$ is the unavoidable additive cost for the entire experiment.  The second objective function is defined as 
\beanno
\Phi_2 = det \Big ( J_{\beta}({\boldsymbol{\theta}_0})^{-1} K_{\beta}({\boldsymbol{\theta}_0})J_{\beta}({\boldsymbol{\theta}_0})^{-1} \Big)
\eeanno
Therefore the overall optimization problem can be framed as
\beanno
&&\underset{\mathscr{D}  }{minimize}
( \Phi_1, \Phi_2) \\
&& \ \text{subject \ to } \Phi_1 <C_1, \ \Phi_2 < C_2,\  \tau_1 < \tau_2 < \cdots < \tau_K, and \ \tau_K < \tau^{*},
\eeanno
where $C_1$ is the prefixed maximum budget that can be expended and $C_2$ is the pre-defined upper bound of the determinant of the covariance matrix, $\tau^{*}$ is the maximum permitted time length of the experiment.  

\noindent This is a multi-objective optimization (MOO) problem. In MOO there is usually no single solution that is optimal with respect to all objectives.  Consequently there are a set of optimal solutions, known as Pareto optimal solutions.  The Pareto optimal solution refers to a solution, around which there is no way of improving any objective without degrading at least one other objective. Without additional information, all these solutions are equally satisfactory.  In this work, we apply Pareto Genetic Algorithm (GA).  A Pareto GA returns a population with many solutions on the Pareto front which is the set of Pareto optimal solutions. The population is ordered based on dominance. Solution $\mathscr{D}_1$ dominates solution $\mathscr{D}_2,$ if 
\beanno
\Phi_1(\mathscr{D}_1) < \Phi_1(\mathscr{D}_2)  \ \text{and} \ \Phi_2(\mathscr{D}_1) \leq \Phi_2(\mathscr{D}_2) \\
\text{or} \ \Phi_1(\mathscr{D}_1) \leq \Phi_1(\mathscr{D}_2)  \ \text{and} \ \Phi_2(\mathscr{D}_1) < \Phi_2(\mathscr{D}_2).
\eeanno
A solution is non-dominated if no solution can be found that dominates it.  

\noindent In this work, we exploit a version of non-dominated sorting GA called NSGA-II proposed by Deb et. al \cite{Deb:2002}. Along with the determination of level of non-dominance, NSGA-II incorporates the density calculation to maintain a good spread of the solutions in the Pareto optimal set.
The tools and the steps of the algorithm are discussed as follows.

\noindent \textbf{Non-dominance rank:}  Let us define non-dominance rank say $i_{rank}.$  At first stage each solution of the population can be compared with all the other solutions to check it is non-dominated or not. All the solutions which are non-dominated are assigned $i_{rank}=1$ and temporarily removed from the population. Next, from the reduced population, we find all the non-dominated solutions and assign $i_{rank}=2$. The process continues until all the solutions are assigned non-dominance rank. 

\noindent \textbf{Rank based on crowding distance :}  Let us define crowding distance rank say $i_{distance}.$  The crowding-distance reflects the density of solutions surrounding a particular solution in the population.  The crowding-distance computation requires sorting the population according to each objective function value in ascending order of magnitude. Let I be any non-dominated set.
For objective function j, $i_{j.distance}$ for solution $\mathscr{D}_i$ is assigned as follows. \\
$l=|I|$\\
for each $\mathscr{D}_i \in I, i_{j.distance}=0$\\
Sort the solutions of I according to the ascending order of magnitude of objective function j.\\
$I=sort(I, \Phi_j)$.\\
In the sorted I, define, $1_{distance}=l_{distance}=\infty$ \\
for $i=2,\ldots,l-1$ \\
$i_{j.distance}=i_{j.distance} + \frac{\Phi_j( \mathscr{D}_{i+1} )- \Phi_j( \mathscr{D}_{i-1} ) }{\Phi^{max}_j -\Phi^{min}_j }$
where $\Phi^{max}_j (\Phi^{min}_j )$ is the maximum (minimum) value of $\Phi_j$ in I.\\
Based on all the objective functions for any solution $i_{distance}= \sum_{j=1}^{2}i_{j.distance}$

\noindent Based on these $i_{rank}$ and $i_{distance},$ Deb et al. \cite{Deb:2002}, define partial order as $<_{n}$ such that 
\begin{center}
    $\mathscr{D}_{i_1}  <_{n} \mathscr{D}_{i_2}$\\
if $\ {i_1}_{rank} < {i_2}_{rank}$ \\
or ${i_1}_{rank} = {i_2}_{rank}$  and ${i_1}_{distance } > {i_2}_{distance}.$
\end{center}
This indicates a lower non-dominance rank is preferred.  Otherwise if non-dominance ranks are same, solution in lesser crowded region is preferred.

\noindent \textbf{Main Loop:} Initially, a population of size N, say $P_0$ is generated.   
Next through binary tournament selection, crossover, and mutation operations off-spring population say $Q_0$ is generated. Elitism is incorporated through $i_{rank}$ and $i_{distance}$ on the combined population.\\
\noindent At any t-th generation of the algorithm, off-spring $Q_t$ is generated from $P_t.$  Next population $R_t$ is formed where $R_t=P_t \cup Q_t$ which is of size 2N.  In $R_t$ find out all the non-dominated solutions and store them in $F_1.$ If $|F_1| <N,$ next from $R_t-F_1$ find out all the non-dominated solutions and store them in $F_2.$  Here, $|  \cdot |$ denotes size of the set.  If $|F_1 \cup F_2|<N$ we continue the process.  Let $l$ be the minimum integer such that $|F_1 \cup F_2 \cup \ldots \cup F_{l-1} \cup F_l| >N.$  Then, we sort the solutions of $F_l$ through the operator $<_{n}$ and choose the best $M=N-|F_1 \cup F_2 \cup \ldots \cup F_{l-1} |$ solutions.  Based on the elitism, thus the next generation will be formed as $P_{t+1}=F_1 \cup F_2 \cup  F_{l-1} \cup \{ \text{best} \ M \ \text{solutions of} \  F_l \}.$

\noindent \textbf{Binary Tournament selection: } In this algorithm, binary tournament selection is executed through the operator $<_{n}.$  In presence of constraints, a solution can be feasible or infeasible.  When two solutions are feasible, they can be compared with the partial order $<_{n}.$  For one feasible and other one infeasible  solution, the feasible one will be chosen.  In presence of two infeasible solutions, the solution having small constraint violation will be chosen.
In this context, the main constraint is set as $\tau_1 \leq \ldots \leq \tau_{K}.$  
For an infeasible solution this inequality will not be satisfied and the constraint violation is measured through $\sum_{i=1}^{K}\sum_{i_j=1}^{i-1} \delta(\tau_{i}-\tau_{i_j} ).$ where,
$\delta(u)=
\begin{cases}
1 \ \text{if} \ u<0\\
0  \quad \text{otherwise}.
\end{cases}$

\noindent \textbf{Crossover:} In this algorithm simulated binary crossover is exploited. 
The steps are given as follows.\\
Select two parents say $\mathscr{D}_{l_1}$ and $\mathscr{D}_{l_2}.$\\
Generate a random number $u \sim U(0,1).$ If $u<P_c,$ generate $r\sim U(0,1)$\\
Compute $\beta$ such that $\beta=
\begin{cases}
(2r)^{\frac{1}{\eta_c + 1}} \quad \text{if} \  r \leq 0.5\\
\Big(\frac{1}{2(1-r)} \Big)^{ \frac{1}{\eta_c + 1}} \quad \text{otherwise}\\
\end{cases}
$\\
where $P_c$ is the crossover probability and $\eta_c$ is called the distribution index. Large 
$\eta_c$ tends to generate children closer to the parents and small $\eta_c$ allows the children to be far from the parents.\\
Generated off-springs are\\
$\mathscr{D}^{new}_{l_1} = 0.5 [ (1+\beta)\mathscr{D}_{l_1} + (1-\beta) \mathscr{D}_{l_1} ],$\\
$\mathscr{D}^{new}_{l_2} = 0.5 [ (1-\beta)\mathscr{D}_{l_1} + (1+\beta) \mathscr{D}_{l_1} ].$\\
For detail study reader may refer to Deb and  Agrawal \cite{Deb:1995}.

\noindent \textbf{Mutation :} A polynomial mutation is implemented in the mutation operation.
For a solution $\mathscr{D}=(\tau_1, \ldots, \tau_{k} ),$ the mutation is operated as follows.\\
Set i==1.\\
Generate $u\sim U(0,1).$
If $u<P_m$
set, $\delta= \frac{min(\tau^{upper}_i - \tau_i, \tau_i - \tau^{lower}_i ) }{ \tau^{upper}_i  - \tau^{lower}_i  }.$ \\
Generate $r\sim U(0,1).$\\
Compute $\delta_q=
\begin{cases}
[2r + (1-2r)(1-\delta)^{(\eta_m +1) } ]^{(\frac{1}{\eta_m + 1}) }  -1 \quad \text{if} \  r \leq 0.5\\
1- [ (2(1-r) + 2(r-0.5)(1-\delta)^{\eta_m+1}]^{\frac{1}{\eta_m + 1} } \quad \text{otherwise}
\end{cases}$\\
$\tau_i=\tau_i + \delta_q(\tau^{upper}_i - \tau^{lower}_i).$\\
Until i++ ==K.

\noindent Here $P_m$ is the mutation probability and $\tau^{upper}_i (\tau^{lower}_i ) $ is the pre-defined upper (lower) bound of $\tau_i.$
For detailed study, the references are Deb et al \cite{Deb:2002}, Hamdan \cite{Ham:2010}.

\section{ \sc Numerical experiment and real data analysis :}
\subsection{\sc Numerical experiment :}
In this section a simulation study has been conducted using the Markov Chain Monte Carlo (MCMC) simulation based on 1000 generations to asses the performances of the developed methods.
 Under two dependent competing causes of failure following MOBE life time model, 20 subjects of interest are put on life testing experiment.  Across different inspection time intervals taken as (0,0.2], (0.2,0.3], (0.3,0.4], the number of failures due to cause 1, cause 2 and both causes are recorded.   
The three different sets of model parameters are taken for the study and these model parameters are contaminated to study the robustness of DPDEs.  Those sets are given in the Table \eqref{tab1}.
\begin{table}[H]
	\caption{Model Parameters (Pure and Contaminated Data)}
	\label{tab1}
	\begin{center}		
		\begin{tabular}{|c|ccc|c|ccc|} 
			\hline
	\multirow{2}{*}{\textbf{S.No.}}	&\multicolumn{3}{c|}{\textbf{Pure Data}} &\multirow{2}{*}{\textbf{S.No.}}&\multicolumn{3}{c|}{\textbf{Contaminated Data}}   \\ 
		 &$\bm{\lambda_0}$&$\bm{\lambda_1}$&$\bm{\lambda_2}$ & &$\bm{\Tilde{\lambda}_0}$&$\bm{\Tilde{\lambda}_1}$&$\bm{\Tilde{\lambda}_2}$\\ 
		\hline
	$\bm{\theta}_1$&4.5 &2.5 &3.5 &$\Tilde{\bm{\theta}}_1$&$\lambda_0$-0.5 &$\lambda_1$-0.6 &$\lambda_2$-0.4 \\
				$\bm{\theta}_2$&6.3 &2.1 &4.2 &	$\Tilde{\bm{\theta}}_2$&$\lambda_0$-0.8 &$\lambda_1$-0.5 &$\lambda_2$-0.6 \\
				$\bm{\theta}_3$&2.0 &3.0 &4.0 &$\Tilde{\bm{\theta}}_3$&$\lambda_0$-0.2&$\lambda_1$-0.1 &$\lambda_2$-0.3 \\
			\hline

		\end{tabular}
	\end{center}
\end{table}
\noindent To obtain DPDEs and MLEs, Coordinate-Descent method is implemented using following steps.
\begin{itemize}
        \item Start iteration process with the initial values $\bm{\theta}^{(0)}=(\lambda_0^{(0)}, \lambda_1^{(0)}, \lambda_2^{(0)})$ where at the $m+1^{th}$ iteration, the estimate of the parameters can be derived as,
        \begin{flalign*}
        \lambda_0^{(m+1)}&=\lambda_0^{(m)}-h\frac{\partial H(\lambda_0^{(m)}, \lambda_1^{(m)}, \lambda_2^{(m)})}{\partial \lambda_0}\\
         \lambda_1^{(m+1)}&=\lambda_1^{(m)}-h\frac{\partial H(\lambda_0^{(m+1)}, \lambda_1^{(m)}, \lambda_2^{(m)})}{\partial \lambda_1}\\
         \lambda_2^{(m+1)}&=\lambda_2^{(m)}-h\frac{\partial H(\lambda_0^{(m+1)}, \lambda_1^{(m+1)}, \lambda_2^{(m)})}{\partial \lambda_2}\\
        &&
        \end{flalign*}
where $H=-l(\bm{\theta})$ for MLEs and $H=H_n(\beta)$ for DPDEs and h is the learning rate taken here as $h=0.01$.
\item The process continues until \{($max\vert \bm{\theta}_j^{(m+1)}-\bm{\theta}_j^{(m)}\vert\,,\,max\vert H(\bm{\theta}^{(m+1)})-H(\bm{\theta}^{(m)})\vert\,;j=0,1,2)<c$\} where $c$ is the threshold value chosen here as $0.0001$.
\end{itemize}
\noindent The Bias of MLEs and DPDEs are given in the Table \eqref{tab2} for the pure data and contaminated data scheme.  It is observed that MLEs are highly affected by contamination as biases of MLEs are increased in contaminated data setting compared to pure data setting.  But observing the bias of the DPDEs, it is evident that the DPDEs are unaffected by the contamination.  The overall behaviour of the bias is that if tuning parameter $\beta$ increases, bias decreases.  Therefore, higher value of $\beta\,;\,0\leq\beta\leq 1$ is preferred for robustness of the DPDEs.

\begin{table}[htb!]
	\caption{Bias of MLE and DPDE (Pure Data and Contaminated Data)}
	\label{tab2}
	\begin{center}		
		\begin{tabular}{|c|ccc|ccc|} 
			\hline
	\multirow{2}{*}{$\bm{\theta_1}$}	&\multicolumn{3}{c|}{\textbf{Pure Data}} &\multicolumn{3}{c|}{\textbf{Contaminated Data}}   \\ \cline{2-7}
		 &$\bm{\lambda_0}$&$\bm{\lambda_1}$&$\bm{\lambda_2}$ &$\bm{\lambda_0}$&$\bm{\lambda_1}$&$\bm{\lambda_2}$\\ 
		\hline
\textbf{MLE} & -0.00055843 &-0.00064128 &-0.00082983 &-0.01949653 &-0.01623247 & -0.01194419 \\
$\bm{\beta=0.2}$ & -0.00031438&-0.00091560 &-0.00086597 &-0.00091536 &-0.00013561 & -0.00052607 \\
$\bm{\beta=0.4}$ &0.00000369 &-0.00001053 &0.00001073 & -0.00012021 &-0.00002257 &-0.00000226  \\
$\bm{\beta=0.6}$ &-0.00002044 &0.00000736 &0.00001134 & -0.00007559 &-0.00001391 &0.00000589  \\
$\bm{\beta=0.8}$ &0.00000448 &-0.00000914 & 0.00000694&-0.00005951 & 0.00000260 &0.00000970  \\
$\bm{\beta=1.0}$ &-0.00000273 & 0.00000454 &0.00000358 &-0.00004392& 0.00000707 & 0.00000160 \\
		\hline
		\multirow{2}{*}{$\bm{\theta_2}$}	&\multicolumn{3}{c|}{\textbf{Pure Data}} &\multicolumn{3}{c|}{\textbf{Contaminated Data}}   \\ \cline{2-7}
		 &$\bm{\lambda_0}$&$\bm{\lambda_1}$&$\bm{\lambda_2}$ &$\bm{\lambda_0}$&$\bm{\lambda_1}$&$\bm{\lambda_2}$\\ 
		\hline
\textbf{MLE} &-0.00087963 &-0.00073572 &-0.00041731 &-0.01227007 &-0.01675230 &-0.01325183  \\
$\bm{\beta=0.2}$ & -0.00001251 &-0.00001281 &0.00001422 &-0.00010689 &-0.00004566 &-0.00003516  \\
$\bm{\beta=0.4}$ &0.00001188 &-0.00001086 &0.00000451 &-0.00007924 &-0.00001869 &-0.00001065  \\
$\bm{\beta=0.6}$ & 0.00001552&0.00000194 &-0.00000558 &-0.00005309 &-0.00000735 &-0.00000435  \\
$\bm{\beta=0.8}$ &-0.00000394 & -0.00000406&0.00000857 &-0.00003254 &-0.00000457 &-0.00000088  \\
$\bm{\beta=1.0}$ &-0.00000407 &0.00000138 & 0.00000177 &-0.00002687 &-0.00000164 & 0.00000051 \\
	\hline
		\multirow{2}{*}{$\bm{\theta_3}$}	&\multicolumn{3}{c|}{\textbf{Pure Data}} &\multicolumn{3}{c|}{\textbf{Contaminated Data}}   \\ \cline{2-7}
		 &$\bm{\lambda_0}$&$\bm{\lambda_1}$&$\bm{\lambda_2}$ &$\bm{\lambda_0}$&$\bm{\lambda_1}$&$\bm{\lambda_2}$\\ 
		\hline
\textbf{MLE} &0.00045100 &-0.00453577 &-0.00632121 &-0.01077694 &-0.01012153 &-0.01448080  \\
$\bm{\beta=0.2}$ &-0.00041102 &-0.00003321 &-0.00025326 &-0.00019593 &-0.00008603 & -0.00038122 \\
$\bm{\beta=0.4}$ & -0.00001909 &0.00001613 &-0.00000079 &0.00001443 & -0.00003119&-0.00004261  \\
$\bm{\beta=0.6}$ &0.00000646 &0.00000058 &0.00000144 & 0.00000708 &-0.00000077 &-0.00004087  \\
$\bm{\beta=0.8}$ &0.00000573 &0.00001019 &-0.00000706 &0.00000836 &-0.00000252 & -0.00001729 \\
$\bm{\beta=1.0}$ &0.00000838 &0.00000539 &-0.00000775 &0.00000665 &0.00000303 & -0.00002132 \\
	\hline
		\end{tabular}
	\end{center}
\end{table}

\noindent  The approximated power of the Wald type test using various set of parameters are calculated which are given in the Table \eqref{tab7}.  From this table it can be observed that as the difference between $\lambda_1$ and $\lambda_2$ increases, power of the test  increases.  It is also observed that as the value of tuning parameter increases, power of the test gradually decreases.

\begin{table}[H]
	\caption{Power of the test}
	\label{tab7}
	\begin{center}		
		\begin{tabular}{|ccc|ccccc|} 
			\hline
\multicolumn{3}{|c|}{\textbf{Parameters}}  &\multicolumn{5}{c|}{\textbf{Power}}\\
\hline
$\bm{\lambda_0}$ &$\bm{\lambda_1}$ &$\bm{\lambda_2}$ & $\bm{\beta=0.2}$ &$\bm{\beta=0.4}$ &$\bm{\beta=0.6}$ &$\bm{\beta=0.8}$ &$\bm{\beta=1.0}$ \\
		\hline
4.5 & 2.5 & 3.0 &0.6024 &0.6012 &0.5999 &0.5983 &0.5967 \\
6.3 & 2.0 & 3.5 &0.7405 &0.7376 &0.7345 &0.7315 & 0.7286 \\
4.5 & 2.5 & 4.0 &0.7634 &0.7615 &0.7591 &0.7563 & 0.7533  \\
4.5 & 2.5 & 5.5 &0.8969 &0.8959 &0.8942 & 0.8922& 0.8899  \\
6.3 & 2.0 & 5.5 &0.9025 &0.9004 &0.8979 & 0.8953& 0.8927  \\
\hline
		\end{tabular}
	\end{center}
\end{table}

\noindent To obtain the optimum inspection time points which would determine the best experimental set-up based on objectives discussed in Section 6, we have set $n=20,$ $P_c=0.9, \eta_c=20,$ $P_m=\frac{1}{K}$, K=3 and $\eta_m=20$.  The parameter values are chosen as $\lambda_0=0.15, \lambda_1=0.02, \lambda_2=0.07$ with tuning parameter $\beta=0.5$ and the population size is set as 50. Here, we set $\tau^{upper}_i=70$ and $\tau^{lower}_i=0$ for $i=1,2,3.$  The Pareto optimal solutions for the initial population and after $100$ generations are given in the Table \eqref{tab3}.  For the initial population, $4$ Pareto optimal solutions are obtained where for the $100^{th}$ generation, number Pareto optimal solutions is increased to $42$.   
\begin{table}[htb!]
	\caption{Optimal Time Points}
	\label{tab3}
	\begin{center}		
		\begin{tabular}{|c|ccc|c|ccc|} 
			\hline
		\multicolumn{8}{|c|}{\textbf{Pareto front of Initial Population}}\\\hline
	\multirow{2}{*}{\textbf{S.No.}}	&\multicolumn{3}{c|}{\textbf{Optimal Time Points}} &\multirow{2}{*}{\textbf{S.No.}}&\multicolumn{3}{c|}{\textbf{Optimal Time Points}}   \\ 
		 &$\bm{\tau_1}$&$\bm{\tau_2}$&$\bm{\tau_3}$ & &$\bm{\tau_1}$&$\bm{\tau_2}$&$\bm{\tau_3}$\\ 
		\hline
1 & 6.428751 & 42.670939 & 65.692024 &  2 & 9.211211 & 15.385411 &  29.028185 \\
3 & 5.390008 & 19.465359 & 57.243752 & 4 & 8.516763 & 34.996492 & 41.205904\\
			\hline
	\multicolumn{8}{|c|}{\textbf{Pareto front of $\bm{100^{th}}$ Generation}}\\\hline
1 & 8.037446 & 19.707918 & 19.817200 & 2 &  4.932801 & 50.605646 & 50.634229 \\
3 & 4.919766 & 45.716619 & 45.806510 & 4 & 5.369326 & 39.244768 & 39.575436 \\
5 & 4.932801 & 50.605646 & 50.634229 & 6 & 4.937966 & 42.961247 & 43.015458 \\
7 & 4.932801 & 50.605646 & 50.634229 & 8 & 4.926927 & 45.608204 & 45.702786 \\
9 & 4.906357 & 45.962833 & 46.043385 & 10 & 4.934266 & 50.862805 & 50.887417 \\
11 & 8.036706 & 19.713485 & 19.822803 & 12 & 4.948874 & 44.928987 & 45.045212 \\
13 & 8.056663 & 19.513107 & 19.015796 & 14 & 4.915135 & 42.830166 & 42.884817 \\
15 & 6.232382 & 36.185958 & 37.218926 & 16 & 4.919698 & 45.717531 & 45.807419 \\
17 & 4.932801 & 50.605646 & 50.634229 & 18 & 4.917877 & 45.762200 & 45.850118 \\
19 & 4.922787 & 43.059069 & 43.106004 & 20 & 4.932719 & 50.584295 & 50.613529 \\
21 & 4.93795 & 42.96136 & 43.01554 & 22 & 4.927139 & 43.066642 & 43.113691 \\
23 & 4.934322 & 50.882763 & 50.907122 & 24 & 4.92063 & 45.74927 & 45.83648 \\
25 & 4.948874 & 44.928987 & 45.045212 & 26 & 8.036706 & 19.713485 & 19.822803 \\
27 & 6.205449 & 37.715003 & 37.990882 & 28 & 4.933978 & 50.637504 & 50.665129 \\
29 & 4.927668 & 45.853462 & 45.944780 & 30 & 4.928646 & 45.570939 & 45.666709 \\
31 & 4.972067 & 50.192955 & 50.223205 & 32 & 4.925974 & 43.038753 & 43.089757 \\
33 & 6.284095 & 35.913353 & 36.985382 & 34 & 4.933086 & 43.027484 & 43.082879 \\
35 & 6.22404 & 36.11990 & 37.90714 & 36 & 4.937925 & 43.022434 & 43.076440 \\
37 & 5.071818 & 42.023778 & 42.166894 & 38 & 4.918082 & 45.757510 & 45.845652 \\
39 & 5.02793 & 42.07590 & 42.34212 & 40 & 4.932968 & 50.947484 & 50.969960 \\
41 & 5.069286 & 42.054582 & 42.301416 & 42 & 6.189256 & 37.782024 & 38.055071 \\
\hline
		\end{tabular}
	\end{center}
\end{table}

\subsection{\sc Real Data Analysis : } 
For the real life implementation of the results obtained in the previous sections, an analysis has been performed on the bivariate data taken from Ebrahimi \cite{Eb:1987}.  For the inspection time intervals (0,0.032], (0.032,0.12], (0.12,0.23],  failure time data due to cause 1, cause 2 and both causes are recorded for first 30 observations from the data set found in Ebrahimi \cite{Eb:1987}.  The failure time values are divided by 10 for the ease of computation.  The description of failure time data is given in the Table \eqref{tab4}.  Failure due to cause 1, cause 2 and both causes are indicated as (1,2,0), respectively. 

\begin{table}[htb!]
	\caption{Failure Time Data}
	\label{tab4}
	\begin{center}		
		\begin{tabular}{|cc|cc|cc|} 
			\hline
	\textbf{Failure Time}&\textbf{Cause}&\textbf{Failure Time}&\textbf{Cause}&\textbf{Failure Time}&\textbf{Cause}\\
		\hline
	0.610	 &1 &0.150  &2 &0.170  &0 \\
	0.017	 &2 &0.180  &0 &0.034  &1 \\
	0.105	 & 1 &0.042 &2 & 0.030 &2 \\
	 0.223	 & 1 &0.250&2 &  0.130 & 0 \\
	 0.397	 &0 & 0.010  &1 & 0.080  &2 \\
	0.047	  &1 & 0.036  & 1 &0.080   &0 \\
	0.004	  &0 &0.006  &2 &0.250   &2 \\
	0.016	 &0 & 0.070 &2 & 0.092  & 2 \\
	 0.046	  &0 &0.030  &2 &0.027  &0 \\
	0.047	  &1 & 0.002 &1 & 0.106 & 2 \\
			\hline
		\end{tabular}
	\end{center}
\end{table}

\noindent To check whether Marshall-Olkin Bivariate Exponential distribution fits the data, a bootstrap based testing has been conducted. The test statistic is defined as  $S=(\sum_{i=1}^{K}\vert N_{ij}-E_{ij}\vert+\vert N_s-E_s\vert$) where $N_{ij}$'s are the number of observed failures and $E_{ij}$'s are the number of expected failures in time interval $( \tau_{i-1}, \tau_i]$ due to cause j for $i=1,\ldots, K$ and $j=0,1,2.$  $N_s$ and $E_s$ respectively, are the number of observed and expected survived units at the time point $\tau_K$.  The MLE or DPDE method can be used to estimate $E_{ij}=n\times p_{ij}\,;j=0,1,2$ and $E_s=n\times p_s.$  

\noindent The MLEs and DPDEs of the model parameters based on the real data set are given in Table \eqref{tab6}. Based on those estimated values of the model parameters, 10000 bootstrap samples are generated and in each bootstrap sample, we compute the test statistic.  The count of the bootstrapped test statistics greater than the real data based test statistic $S$ divided by the number of bootstrap sample is the approximate p-value.  The values of the real data based test statistics and the corresponding approximate p-values using MLE and DPDE for different tuning parameters ($\beta$) are given in Table \eqref{tab5}.  The obtained approximated p-values indicate that MOBE can be applied as the life time distribution for this real data set. 

\begin{table}[H]
	\caption{Approximate p-value Calculation}
	\label{tab5}
	\begin{center}		
		\begin{tabular}{|c|c|c|} 
			\hline
	$\bm{\theta}$&\textbf{Test Statistic}&\textbf{Approximate p-value}\\
		\hline
\textbf{MLE}& 15.03554 &0.2213 \\
$\bm{\beta=0.2}$&15.03560 &0.2084  \\
$\bm{\beta=0.4}$&15.03565 &0.2178 \\
$\bm{\beta=0.6}$&15.03570 &0.2118 \\
$\bm{\beta=0.8}$&15.03573 & 0.2168 \\
$\bm{\beta=1.0}$&15.03576 &0.2235 \\
			\hline
		\end{tabular}
	\end{center}
\end{table}
\noindent  The MLEs and DPDEs are calculated using the Coordinate Descent algorithm.  In this algorithm, the initial values of the model parameters $(\lambda_0=3.5, \lambda_1=1.5, \lambda_2=2.5)$ are obtained through grid-search procedure.  Also bootstrap estimates of the bias (BT Bias) are computed for each of the estimators which are reported in Table \eqref{tab6}.  

\begin{table}[htb!]
	\caption{MLEs and DPDEs and Bootstrap Estimates of Bias}
	\label{tab6}
	\begin{center}		
		\begin{tabular}{|c|cc|cc|cc|} 
			\hline
		\multirow{2}{*}{$\bm{\theta}$}&\multicolumn{2}{c|}{$\bm{\lambda_0}$} &\multicolumn{2}{c|}{$\bm{\lambda_1}$} &\multicolumn{2}{c|}{$\bm{\lambda_2}$}\\\cline{2-7}
&\textbf{Estimate}&\textbf{BT Bias}&\textbf{Estimate} & \textbf{BT Bias}&\textbf{Estimate}&\textbf{BT Bias} \\
\hline
\textbf{MLE} &3.500992 &-0.027849 &1.500634 &-0.017526 &2.499711 &-0.007305 \\
$\bm{\beta=0.2}$ &3.500711 &-0.021686 &1.500497 &-0.012530 &2.499757 &-0.001851 \\
$\bm{\beta=0.4}$ &3.500498 &-0.024127 &1.500381 &-0.013548 &2.499802 &-0.000306 \\
$\bm{\beta=0.6}$ &3.500344 &-0.019039 &1.500287 &-0.008495 &2.499841 & -0.001805\\
$\bm{\beta=0.8}$ &3.500236 &-0.019323 &1.500214 &-0.013260 &2.499874 &-0.000929 \\
$\bm{\beta=1.0}$ &3.500161 &-0.017740 &1.500158 &-0.010565 &2.499902 &-0.003314 \\
			\hline

		\end{tabular}
	\end{center}
\end{table}

\section{Conclusion}
In this work, a robust estimation method has been developed to estimate the life time distribution  under two dependent competing risks which is modelled by Marshall-Olkin bivariate exponential distribution.  The point estimation has been studied through robust minimum density power divergence estimator (MDPDE) and also we have computed maximum likelihood estimators (MLE).  Testing of hypothesis has been performed through Wald type test statistic based on asymptotic distribution of MDPDE.  It is observed through simulation study that MLEs provide misleading results in presence of contamination while MDPDEs remain unaffected by contamination.  The influence function of the MDPDE and the test statistic have also been derived which measures the robustness analytically.  In the study of power of wald type test, it is observed that for $H_0:\lambda_1=\lambda_2$, power would be high if difference between $\lambda_1$ and $\lambda_2$ is high  and value of tuning parameter is small.  In determination of optimal inspection times,
Pareto Genetic Algorithm has been applied and it is observed that by increasing the number of generations, number of Pareto optimal solutions are also increased.  Real data analysis has also been conducted to asses the performances of the theoretical results in practical situations.

\noindent The model analysed here can be studied by incorporating covariates and can also be extended to the analysis of missing information on covariates, masked cause of failures etc.  Same study can be conducted taking other lifetime distributions and model can be applied on some particular situations like reliability analysis of one-shot devices.  Efforts in this direction is under way and we would report these findings as soon as possible.

\subsection*{Conflict of interest}
The authors declare that they have no conflict of interest.

\section*{Appendix:}

Proof of \textbf{Result 1:}\\
Here, true value of ${ \boldsymbol{\theta}}$ is denoted by ${ \boldsymbol{\theta}_0 }=(\lambda_{00}, \lambda_{10}, \lambda_{20})^T$ and MDPDE ${\what{\boldsymbol{\theta}} }_{\beta}=(\lambda_{\beta 0}, \lambda_{\beta 1}, \lambda_{\beta 2})^T.$\\
Define $M=3*K+1$ and\\
$p_l=p_{3(i-1)+j+1}$ and $N_l=N_{3(i-1)+j+1}$ for $j=0,1,2$ and $i=1, \ldots, M-1$,\
$p_{M}=p_s,$ $N_{M}=N_s.$ \\
Therefore, $H_{n}(\beta)$ can be expressed as
\beanno
H_{n\beta}( \boldsymbol{\theta}  )= \sum_{l=1}^{M}p^{\beta+1}_l - \frac{1+\beta}{\beta}\sum_{l=1}^{M}\left(\frac{N_{l}}{n}p_{l}^{\beta}\right).
\eeanno
Define, $X_s=(X_{s1}, X_{s2}, \ldots, X_{sM} )\sim Multinomial(1, p_1, p_2, \ldots, p_M ).$
Therefore $N_l$ can be expressed as $N_l= \sum_{s=1}^{n}X_{sl}$ and $H_{n}(\beta)$ can be re-written as
\beanno
H_n(\beta)& =& \frac{1}{n}  \sum_{s=1}^{n} \Big( \sum_{l=1}^{M} p^{\beta +1}_l  - \frac{1+\beta}{\beta}\sum_{l=1}^{M} X_{sl} p_{l}^{\beta} \Big)\\
&=& \frac{1}{n}  \sum_{s=1}^{n} V_{\beta}(X_s, \boldsymbol{\theta} )
\eeanno
where, $V_{\beta}(X_s, \boldsymbol{\theta} )= \sum_{l=1}^{M} p^{\beta +1}_l  - \frac{1+\beta}{\beta}\sum_{l=1}^{M} X_{sl} p_{l}^{\beta}.$ \\
Denote, $H_{nj\beta}= \frac{\partial H_n(\beta)}{\partial \lambda_j}=\frac{1}{n}  \sum_{s=1}^{n} \frac{\partial V_{\beta}(X_s, \boldsymbol{\theta} )}{\partial \lambda_j}$ for $j=0,1,2.$
Here, we get $E( \frac{\partial V_{\beta}(X_s, \boldsymbol{\theta} )}{\partial \lambda_j} )=0$
and
\beanno
 Var\Big(\frac{\partial V_{\beta}(X_s,
 \boldsymbol{\theta} )}{\partial \lambda_j} \Big)
&=&(\beta+1)^2 Var\Big( \sum_{l=1}^{M}X_{sl} p^{\beta -1}_l \frac{\partial p_l}{\partial \lambda_j} \Big)\\
&=&(\beta+1)^2 \Bigg( \sum_{l=1}^{M}  p^{2(\beta -1)}_l  p_l(1-p_l) (\frac{\partial p_l}{\partial \lambda_j})^2 - 2 \sum\limits_{\substack{1 \leq l_1<l_2 \leq M}} p^{(\beta -1)}_{l_1}  p^{(\beta -1)}_{l_2} p_{l_1} p_{l_2}      \frac{\partial p_{l_1} }{\partial \lambda_j}\frac{\partial p_{l_2}}{\partial \lambda_j} \Bigg)
\eeanno
for $j=0,1,2.$
\beanno
Cov(\frac{\partial V_{\beta}(X_s, \boldsymbol{\theta} )}{\partial \lambda_{j_1} }, \frac{\partial V_{\beta}(X_s, \boldsymbol{\theta} )}{\partial \lambda_{j_2} })
&=& (\beta+1)^2 \Bigg(  \sum_{l=1}^{M}  p^{2(\beta -1)}_l  p_l(1-p_l) 
\frac{\partial p_l}{\partial \lambda_{j_1} } \frac{\partial p_l}{\partial \lambda_{j_2} }   \\
&& \ -2 \sum\limits_{\substack{1 \leq l_1<l_2 \leq M}} p^{(\beta -1)}_{l_1}  p^{(\beta -1)}_{l_2} p_{l_1} p_{l_2}      \frac{\partial p_{l_1} }{\partial \lambda_{j_1} }\frac{\partial p_{l_2}}{\partial \lambda_{j_2} } \Bigg)
\eeanno
for $j_1, j_2= 0,1,2$ and $j_1 \neq j_2.$\\
Define matrix, $K_{\beta}(\boldsymbol{\theta} )$ where
$K_{\beta}(\boldsymbol{\theta} )_{jj} = \frac{1}{(\beta +1)^2} Var(\frac{\partial V_{\beta}(X_s, \boldsymbol{\theta} )}{\partial \lambda_j})$
and\\
$K_{\beta}(\boldsymbol{\theta} )_{j_1 j_2}= \frac{1}{(\beta +1)^2} Cov(\frac{\partial V_{\beta}(X_s, \boldsymbol{\theta} )}{\partial \lambda_{j_1} }, \frac{\partial V_{\beta}(X_s, \boldsymbol{\theta} )}{\partial \lambda_{j_2} })$
for $j=0,1,2$ and $j_1,j_2=0,1,2; j_1 \neq j_2.$

\noindent Define, 
$T_{n\beta}=(T_{0n\beta}, T_{1n\beta},T_{2n\beta} )$ where,
\beanno
T_{0n\beta} &=&-\sqrt{n} H_{n0\beta} (\boldsymbol{\theta}_0)=-\sqrt{n} \frac{\partial H_{n 0 \beta} (\boldsymbol{\theta})   }{\partial \lambda_0 }|  \boldsymbol{\theta}=\boldsymbol{\theta}_0,\\
T_{1n\beta} &=& -\sqrt{n} H_{n1\beta} (\boldsymbol{\theta}_0)=-\sqrt{n} \frac{\partial H_{n 1 \beta} (\boldsymbol{\theta})   }{\partial \lambda_1 }|  \boldsymbol{\theta}=\boldsymbol{\theta}_0,\\
T_{2n\beta} &=& -\sqrt{n} H_{n2\beta} (\boldsymbol{\theta}_0)=-\sqrt{n} \frac{\partial H_{n 2 \beta} (\boldsymbol{\theta})   }{\partial \lambda_2 }|  \boldsymbol{\theta}=\boldsymbol{\theta}_0.
\eeanno

\noindent Applying Central Limit Theorem, $T_{n\beta} \sim N(0_3, (\beta+1)^2 K_{\beta}(\boldsymbol{\theta}_0 ) ).$

\noindent Next, we get,
 \beanno
 \frac{\partial H_{n j_1\beta} }{\partial \lambda_{j_2} }=\frac{ \partial^2 V_{\beta}(X_s, \boldsymbol{\theta} ) }{\partial \lambda_{j_1} \partial \lambda_{j_2} }  &=&\sum_{l=1}^{M} \Big [ (\beta+1) \beta p^{\beta -1}_l \frac{\partial p_l}{\partial \lambda_{j_1} } \frac{\partial p_l}{\partial \lambda_{j_2} }  + (\beta+1) p^{\beta }_l  \frac{\partial ^2 p_l}{\partial \lambda_{j_1}   \partial \lambda_{j_2}  } \Big ]\\
& & - \sum_{l=1}^{M} \Big[  (\beta+1)(\beta-1) X_{sl}  p^{\beta -2}_l \frac{\partial p_l}{\partial \lambda_{j_1} } \frac{\partial p_l}{\partial \lambda_{j_2} }
 +(\beta+1)X_{sl}  p^{\beta-1 }_l  \frac{\partial ^2 p_l}{\partial \lambda_{j_1}   \partial \lambda_{j_2}  } \Big]
\eeanno
and $\frac{1}{n} \sum_{s=1}^{n}X_{sl} \xrightarrow[\text{}]{\text{ P}} p_l.$

\noindent Therefore, it is evident that, 
$\frac{\partial H_{n j_1\beta} }{\partial \lambda_{j_2} } \xrightarrow[\text{}]{\text{ P}} (\beta+1) \sum_{l=1}^{M} \Big ( p^{\beta -1}_l \frac{\partial p_l}{\partial \lambda_{j_1} } \frac{\partial p_l}{\partial \lambda_{j_2} } \Big ).
$

\noindent Taylor series expansion of $H_{nj\beta}( \boldsymbol{\theta} )$ around $\boldsymbol{\theta}_0 $ gives
\beanno
H_{nj\beta}( \boldsymbol{\theta} ) &=& H_{n j \beta}( \boldsymbol{\theta}_0 )
+ \sum_{k=0}^{2} \frac{\partial H_{nj\beta} (\boldsymbol{\theta}  ) }{\partial \lambda_k} |_{ \boldsymbol{\theta}=\boldsymbol{\theta}_0 } (\lambda_k - \lambda_{k0} ) \\ && + \frac{1}{2} \sum_{j_1=0}^{2} \sum_{j_2=0}^{2} 
\frac{\partial^2 H_{nj\beta} (\boldsymbol{\theta}  ) }{\partial \lambda_{j_1}  \partial \lambda_{j_2} } |_{ \boldsymbol{\theta}=\boldsymbol{\theta}_0 } (\lambda_{j_1} - \lambda_{j_1 0} )(\lambda_{j_2} - \lambda_{j_2 0} ).
\eeanno
As, $H_{nj\beta}( {\what{\boldsymbol{\theta}} }_{\beta}) =0,$ it can be written that,
\beanno
-\sqrt{n} H_{nj\beta}( \boldsymbol{\theta}_0 )= \sqrt{n} \sum_{k=0}^{2} \Big [  \frac{\partial H_{nj\beta} (\boldsymbol{\theta}  ) }{\partial \lambda_k} |_{ \boldsymbol{\theta}=\boldsymbol{\theta}_0 } 
+ \frac{1}{2}\sum_{j_1=0}^{2} 
\frac{\partial^2 H_{n j \beta} (\boldsymbol{\theta}  ) }{\partial \lambda_{k}  \partial \lambda_{j_1} } |_{ \boldsymbol{\theta}=\boldsymbol{\theta}_0 } (\what{\lambda}_{\beta j_1} - \lambda_{j_1 0} )\Big] (\hat{\lambda}_{\beta k} - \lambda_{k 0} ).
\eeanno

\noindent Define, $A_{jkn\beta} = \frac{\partial H_{nj\beta} (\boldsymbol{\theta}  ) }{\partial \lambda_k} |_{ \boldsymbol{\theta}=\boldsymbol{\theta}_0 } 
+ \frac{1}{2}\sum_{j_1=0}^{2} 
\frac{\partial^2 H_{n j \beta} (\boldsymbol{\theta}  ) }{\partial \lambda_{k}  \partial \lambda_{j_1} } |_{ \boldsymbol{\theta}=\boldsymbol{\theta}_0 } (\what{\lambda}_{\beta j_1} - \lambda_{j_1 0} )$ and it is easy to show that
$A_{jkn\beta} \xrightarrow[\text{}]{\text{ P}} (\beta+1) \sum_{l=}^{M} p^{\beta-1}_l \frac{\partial p_l}{\partial \lambda_{j} } \frac{\partial p_l}{\partial \lambda_{k} }.$\\
\noindent Define, 3x3 matrix $A_{n\beta}$ with $j,k$th element $ A_{jkn \beta  }. $  $A_{n\beta} \xrightarrow[\text{}]{\text{ P}}  (\beta+1) J_{\beta}(\boldsymbol{\theta}_0 )$
where 
\bea
\label{mt}
J_{\beta}(\boldsymbol{\theta}_0 )= ((  \sum_{l=1}^{M}  p^{\beta -1}_l \frac{\partial p_l}{\partial \lambda_{j_1} } \frac{\partial p_l}{\partial \lambda_{j_2} } )_{j_1,j_2} ).
\eea

\noindent Define, $Z_{kn \beta} =\sqrt{n} ( \what{\lambda}_{\beta k} - \lambda_{k0} ),$ for $k=0,1,2.$
Therefore, $T_{jn\beta}$ can be expressed as $T_{jn\beta}=\sum_{k=0}^{2}A_{jkn \beta} Z_{kn \beta}.$  Denote, $Z_{n\beta}=( Z_{0n \beta}, Z_{1n \beta}, Z_{2n \beta} ) $ and therefore, we obtain, $T_{n\beta}=A_{n\beta} Z_{n\beta}.$\\
It can be expressed that $Z_{n\beta}=A^{-1}_{n\beta}T_{n\beta} \implies
\sqrt{n} ( {\what{\boldsymbol{\theta}}}_{\beta}  - \boldsymbol{\theta}_0)=Z_{n\beta} \sim
N(0_3, J^{-1}_{\beta}(\boldsymbol{\theta}_0) K_{\beta}(\boldsymbol{\theta}_0 )  J^{-1}_{\beta}(\boldsymbol{\theta}_0 )   ).$

\noindent Proof of \textbf{Result 2:}\\
In \eqref{if-1}, replacing G by $G_{\epsilon}=(1-\epsilon)F_{\boldsymbol{\theta}}  + \epsilon \Delta_{\boldsymbol{x}},$
differentiating with respect to $\epsilon,$ and taking $\epsilon \to 0,$ we obtain
\beanno
\Big[ A_{\beta}(\boldsymbol{\theta} ) + (\beta+1) J_{\beta}(\boldsymbol{\theta} )\Big ] IF - \Big[ A_{\beta}(\boldsymbol{\theta} ) + \beta J_{\beta}(\boldsymbol{\theta} ) \Big] IF
= \sum_{i=1}^{k} \sum_{j=0}^{2}\Bigg(\int_{I_{ij}} d\Delta_{\boldsymbol{x}} - P_{ij} \Bigg) \frac{\partial \log p_{ij} }{\partial \boldsymbol{\theta}    } \\ + \Bigg(\int_{I_{s}} d\Delta_{\boldsymbol{x}} - P_{s} \Bigg) \frac{\partial \log p_{s} }{\partial \boldsymbol{\theta}    }
\eeanno
\begin{flalign*}
 \implies IF &= J^{-1}_{\beta}(\boldsymbol{\theta}  )\sum_{i=1}^{k} \sum_{j=0}^{2}(\delta_{I_{ij}}(\boldsymbol{x}) - P_{ij} ) \frac{\partial \log p_{ij} }{\partial \boldsymbol{\theta}    }  + (\delta_{I_{s}}(\boldsymbol{x})  - P_{s} ) \frac{\partial \log p_{s} }{\partial \boldsymbol{\theta}    }.
 &&
\end{flalign*}
Here, $A_{\beta}(\boldsymbol{\theta})= ((\sum_{i=1}^{k} \sum_{j=0}^{2} P^{\beta+1}_{ij} \frac{\partial^2 \log p_{ij} }{\partial \theta_{l_2} \partial \theta_{l_1}    } +  P^{\beta+1}_{s} \frac{\partial^2 \log p_{s} }{\partial \theta_{l_2} \partial \theta_{l_1}    }) _{l_1,l_2} )$,
$J_{\beta}(\boldsymbol{\theta}  )$  as defined in \eqref{mt}\\
and 
$\delta_{A}(\boldsymbol{x}) = 
\begin{cases}
1 \quad \text{if} \  \boldsymbol{x} \in A\\
0 \quad \text{otherwise}.\\
\end{cases}
$
\\


\begin{thebibliography}{apalike}

\bibitem{A:2016}
Austin, P.C., Lee, D.S. and Fine, J.P., 2016. Introduction to the analysis of survival data in the presence of competing risks. Circulation, 133(6), pp.601-609.

\bibitem{Bai:2020}
Bai, X., Shi, Y., Liu, Y., \& Zhang, C. (2020). Statistical inference for constant‐stress accelerated life tests with dependent competing risks from Marshall‐Olkin bivariate exponential distribution. Quality and Reliability Engineering International, 36(2), 511-528.

\bibitem{BN:2009} Balakrishnan, N. and Han, D., 2009. Optimal step-stress testing for progressively Type-I censored data from exponential distribution. Journal of statistical planning and inference, 139(5), pp.1782-1798.

\bibitem{BL:2012}
Balakrishnan, N. and Ling, M.H., 2012. EM algorithm for one-shot device testing under the exponential distribution. Computational Statistics \& Data Analysis, 56(3), pp.502-509.

\bibitem{BL1:2014}
Balakrishnan, N. and Ling, M.H., 2014. Gamma lifetimes and one-shot device testing analysis. Reliability Engineering \& System Safety, 126, pp.54-64.

\bibitem{BL2:2019}
Balakrishnan, N., Castilla, E., Martín, N. and Pardo, L., 2019. Robust estimators for one-shot device testing data under gamma lifetime model with an application to a tumor toxicological data. Metrika, 82(8), pp.991-1019.


\bibitem{BN1:2015}
Balakrishnan, N., So, H.Y. and Ling, M.H., 2015. EM algorithm for one-shot device testing with competing risks under exponential distribution. Reliability Engineering \& System Safety, 137, pp.129-140.

\bibitem{BN2:2015}
Balakrishnan, N., So, H.Y. and Ling, M.H., 2015. EM algorithm for one-shot device testing with competing risks under Weibull distribution. IEEE Transactions on Reliability, 65(2), pp.973-991.

\bibitem{BN3:2015}
Balakrishnan, N., So, H.Y. and Ling, M.H., 2015. A Bayesian approach for one-shot device testing with exponential lifetimes under competing risks. IEEE Transactions on Reliability, 65(1), pp.469-485.


\bibitem{Basu:1998} Basu, A., Harris, I.R., Hjort, N.L. and Jones, M.C., 1998. Robust and efficient estimation by minimising a density power divergence. Biometrika, 85(3), pp.549-559.

\bibitem{Basu:2017} Basu A, Ghosh A, Mandal A, Mart´ın N, Pardo L. A Wald-type test statistic for testing linear hypothesis in logistic regression models based on minimum density power divergence estimator. Electronic Journal of Statistics. 2017; 11(2):2741-2772.


\bibitem{BR:2020} Bhattacharya, R., Saha, B.N., Farías, G.G. and Balakrishnan, N., 2020. Multi-criteria-based optimal life-testing plans under hybrid censoring scheme. Test, 29(2), pp.430-453.

\bibitem{Cai:2017} 
Cai, J., Shi, Y., \& Liu, B. (2017). Analysis of incomplete data in the presence of dependent competing risks from Marshall–Olkin bivariate Weibull distribution under progressive hybrid censoring. Communications in Statistics-Theory and Methods, 46(13), 6497-6511.

\bibitem{Cal:2021} Calvino, A., Martin, N. and Pardo, L., 2021. Robustness of Minimum Density Power Divergence Estimators and Wald-type test statistics in loglinear models with multinomial sampling. Journal of Computational and Applied Mathematics, 386, p.113214.

\bibitem{Cr:2001}
Crowder, M.J., 2001. Classical competing risks. Chapman and Hall/CRC.

\bibitem{Deb:2002} Deb K, Pratap A, Agarwal S, Meyarivan TA. A fast and elitist multiobjective genetic algorithm: NSGA-II. IEEE transactions on evolutionary computation. 2002 Aug 7;6(2):182-97.

\bibitem{Deb:1995} Deb, K. and Agrawal, R.B., 1995. Simulated binary crossover for continuous search space. Complex systems, 9(2), pp.115-148.


\bibitem{Deb:1999} Deb, K. and Agrawal, S., 1999. A niched-penalty approach for constraint handling in genetic algorithms. In Artificial neural nets and genetic algorithms (pp. 235-243). Springer, Vienna.


\bibitem{Deb:2014} Deb, K. and Deb, D., 2014. Analysing mutation schemes for real-parameter genetic algorithms. Int. J. Artif. Intell. Soft Comput., 4(1), pp.1-28.


\bibitem{Dutta:2022}
Dutta, S., \& Kayal, S. (2022). Bayesian and non‐Bayesian inference of Weibull lifetime model based on partially observed competing risks data under unified hybrid censoring scheme. Quality and Reliability Engineering International.


\bibitem{Eb:1987} Ebrahimi, N. 1987. Analysis of bivariate accelerated life test data for the bivariate exponential distribution. American Journal of Mathematical and Management Sciences,  7(1-2), 175-190.

\bibitem{Faraz:2013}
Faraz, A., \& Saniga, E. (2013). Multiobjective Genetic Algorithm Approach to the Economic Statistical Design of Control Charts with an Application to bar and S2 Charts. Quality and Reliability Engineering International, 29(3), 407-415.

\bibitem{FH:2015}  
Feizjavadian, S. H., \& Hashemi, R. (2015). Analysis of dependent competing risks in the presence of progressive hybrid censoring using Marshall–Olkin bivariate Weibull distribution. Computational Statistics \& Data Analysis, 82, 19-34.

\bibitem{Ham:2010} Hamdan, M., 2010. On the disruption-level of polynomial mutation for evolutionary multi-objective optimisation algorithms. Computing and Informatics, 29(5), pp.783-800.

\bibitem{KM:2021} 
Kundu, D., \& Mondal, S. (2021). Analyzing competing risks data using bivariate Weibull-geometric distribution. Statistics, 55(2), 276-295.


\bibitem{DK:2022} Kundu, D., 2022. Bivariate Semi-parametric Singular Family of Distributions and its Applications. Sankhya B, pp.1-27.


\bibitem{Ling:2019}
Ling, M.H., 2019. Optimal design of simple step-stress accelerated life tests for one-shot devices under exponential distributions. Probability in the Engineering and Informational Sciences, 33(1), pp.121-135.

\bibitem{Liu:2015}
Liu, X., Zheng, S., Feng, J., \& Chen, T. (2015). Reliability reallocation for fuel cell vehicles based on genetic algorithm. Quality and Reliability Engineering International, 31(8), 1495-1502.

\bibitem{Lyu:2022}
Lyu, H., Qu, H., Ma, L., Wang, S., \& Yang, Z. (2022). Reliability assessment of a system with multi‐shock sources subject to dependent competing failure processes under phase‐type distribution. Quality and Reliability Engineering International.

\bibitem {MO:1967} Marshall, A.W. and Olkin, I., 1967. A multivariate exponential distribution. Journal of the American Statistical Association, 62(317), pp.30-44.


 \bibitem{Ng:2004} 
 Ng, H. K. T., Chan, P. S., \& Balakrishnan, N. (2004). Optimal progressive censoring plans for the Weibull distribution. Technometrics, 46(4), 470-481.

\bibitem{Park:2000}
Parkinson, D. B. (2000). Robust design employing a genetic algorithm. Quality and Reliability Engineering International, 16(3), 201-208.


\bibitem{P:1978}
Prentice, R.L., Kalbfleisch, J.D., Peterson Jr, A.V., Flournoy, N., Farewell, V.T. and Breslow, N.E., 1978. The analysis of failure times in the presence of competing risks. Biometrics, pp.541-554.

\bibitem{SX:2018} 
Shen, Y., \& Xu, A. (2018). On the dependent competing risks using Marshall–Olkin bivariate Weibull model: Parameter estimation with different methods. Communications in Statistics-Theory and Methods, 47(22), 5558-5572.


\bibitem{wang:2020}
Wang, Y. C., Emura, T., Fan, T. H., Lo, S. M., \& Wilke, R. A. (2020). Likelihood‐based inference for a frailty‐copula model based on competing risks failure time data. Quality and Reliability Engineering International, 36(5), 1622-1638.

\bibitem{Yang:2021}
Yang, K., Wang, Y. J., Yao, Y. N., \& Fan, S. D. (2021). Remaining useful life prediction via long‐short time memory neural network with novel partial least squares and genetic algorithm. Quality and Reliability Engineering International, 37(3), 1080-1098.


\end{thebibliography}
\end{document}